# Community-Based Resilience: Digital Technologies for Living within Planetary Boundaries

Catherine Mulligan[*], Giaime Berti [2], and Seema Gadh Kumar[3]

[1]Imperial College London, Institute of Security Science and Technology, London, SW7 2AZ, UK
[2]Scuola Superiore Sant'Anna, School of Advanced Studies, Pisa, Italy
[3]City of Oregon, Chief of Community Technology, Oregon, USA

[*]c.mulligan@imperial.ac.uk

## Abstract

The world faces increasing challenges related to climate change and shifting geopolitical situations. The COVID-19 pandemic clearly illustrated the need for resilience not just within civil engineering but also within social and economic systems. While significant work has focused on applying digital technologies to solve the Sustainable Development Goals (SDGs), less effort has been placed on the ability of digital technologies to enable humanity to continue to live effectively within the Planetary Boundaries (PB). Within this paper, we perform a Systematic Literature Review using the Preferred Reporting Items for Systematic Reviews and Meta-Analyses (PRISMA) methodology. Nine hundred and twenty papers were reviewed. Based on the Systematic Literature Review (SLR), we have outlined specific methods to apply digital technologies to enable humanity to live within PB. Critical capabilities provided by digital technologies - particularly the ability to create dynamic networks for enabling humanity to live within PB - are identified and illustrated in food and agriculture. The paper closes with a brief assessment of the concept's applicability to other critical national infrastructure (CNI).

## Introduction

The world is facing severe challenges from climate change, possible pandemics similar in nature to COVID-19, and geopolitical shifts that are bringing increasing instability and war. As humanity moves forward into the 21st century, it has become evident that the infrastructure implemented for CNI in the so-called global north is often ageing and increasingly poorly designed for the new climate realities that humanity will face. Meanwhile, accessing essential services through CNI remains a complex and challenging issue in the so-called global south.

Innovative solutions, including those enabled by digital technologies, are increasingly needed due to climate change's wide range of severe global impacts. These include rising temperatures contributing to more frequent and intense extreme weather events such as heatwaves, floods, and droughts (IPCC, 2019), which in turn are causing flow-on effects to the delivery of food, water and energy. Coastal regions are particularly vulnerable due to sea-level rise, which leads to increased flooding and saltwater intrusion, threatening freshwater resources and agriculture. Biodiversity is also at risk, as many species struggle to adapt to the rapid changes, leading to ecosystem disruptions. Additionally, climate change affects human health, with more heat-related illnesses and the spread of diseases becoming common.

Vulnerable populations, such as low-income groups and coastal communities, are disproportionately affected. (EUClimateChangeConsequences, 2024).

Climate change (regional or global) has played a role in the collapse or transformation of numerous previous societies (Kemp, 2022). Human societies have often been locally adapted to specific climatic niches; however, over the last century or so, the delivery of food, products and services has been increasingly globalized, reducing the reliance on local production and creating large-scale supply chains. As climate change increases, complex feedback loops are created, possibly damaging infrastructure and in the case of food, the "potential risk of multi-breadbasket failure is increasing" (IPCC, 2019). As a result, the cumulative impacts of warming and biodiversity loss may overwhelm societal adaptive capacity (Kemp, 2022), meaning that research into different methods of delivering the services usually provided by CNI is critical. However, as discussed in our literature review, most research in this domain takes a "business as usual" approach. This paper aims to address this imbalance.

While nations define CNI differently, it can be roughly divided into energy, water, food, transport, telecommunications, and health care. In some cases, it can also include banking and finance. Without these services, nations or regions would struggle to ensure basic living standards for citizens and with prolonged shortages, civil unrest may occur.

Digital technologies have gained an increasing foothold in society, moving far beyond "Information Technology" for use inside companies. Mobile phones, cloud computing, blockchain, Internet of Things (IoT), AI and many other digital solutions have combined to transform how society functions. Much has been written about these transformational effects, including how various technologies may help the world achieve the Sustainable Development Goals (SDGs). Focusing on the SDGs, however, has yet to enable the level of innovation required to solve the issues of increasing climate change (Kim, 2023, Kubiszewski et al.,2021); solutions focus on the maintenance or increased delivery of goods and services in the same large-scale systems that are themselves exposed to the risks and threats. The SDGs can, therefore, look to solve increasingly complex problems with the same techniques as previous generations. This leads us to solve the wrong problems, and the actual capabilities of digital technologies are, therefore, being overlooked. For example, work on SDG 2 has recently returned to the same levels as in 2015 (FAO, 2024a), when the SDGs were defined, meaning that little tangible progress has been made. Our current methods of solving the issues outlined by the SDGs are, therefore, not working. We cannot merely apply digital technologies to deliver the same solutions that do not work; we need to develop new solutions.

Using a PB approach enables us to design new solutions to these issues. Ensuring humanity can live within PB dramatically changes how we apply digital technologies. When approached through a PB lens, we cannot merely apply more of the same solutions; using PB forces us to develop new approaches.

Several nations follow the concept of "Total Defence" (Regeringskansliet, 2024). TD is a defence policy whereby military and civil defence work together to assure the defence of a nation. There are various instances of these types of activities in different countries. However, these programs focus mainly on defence, i.e., preparing for and responding to war. However, in the face of increasing uncertainty related to climate change, creating Total Resilience (TR) plans will become increasingly critical for most nations to develop. Where TD focuses on response to war, TR focuses on the ability of society to continue to operate during more extended periods of instability due to climate events. Such events may occur internally or externally in a nation but significantly impact the freedom of movement and the functioning of society. Digital technologies have a crucial role to play in the delivery of the concepts of TR. This paper contributes to guidance on using digital technologies to assist humanity in living within PB.

## Structure of paper

This paper is split into three sections:

1. A systematic literature review of the work completed related to digital technologies and planetary boundaries
2. An outline of how digital technologies can assist humanity to remain inside PB, using food and agriculture as an example
3. The paper concludes with a section on future research, which outlines how the proposed approach can be applied to other Critical National Infrastructure (CNI) areas such as communications, energy, etc.

## Systematic Literature Review

We have followed the Kitchenham and Charters (Kitchenham, 2007) guide for systematic reviews and used the Preferred Reporting Items for Systematic Reviews and Meta-Analyses (PRISMA) approach for a detailed structured literature analysis (Page, 2021). The Kitchenham report outlines three stages for an SLR – Plan, Conduct and Report, illustrated in Figure 1.

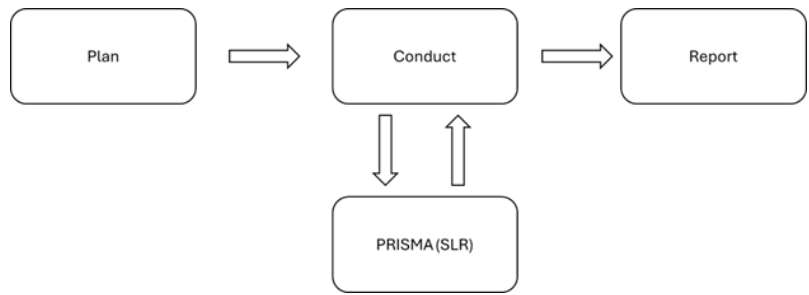

Figure 1: SLR Approach

### Method

We analysed existing review articles in the "Plan" stage. We performed an SLR on digital technologies and their use to ensure humanity can live within PB, focusing on how the connection between the two domains is being approached within the literature. During the "Conduct" stage, due to the large number of papers and the broad nature of the topic in question, a rigorous approach to conducting the SLR was selected to ensure the review was robust, transparent, and replicable. PRISMA is an evidence-based minimum set of items aimed at helping scientific authors report a wide array of systematic reviews and meta-analyses. PRISMA was initially developed within medical research, explicitly aiming to understand the benefits/harms of a healthcare intervention (Page,2021). However, this approach is increasingly used within technical research as a robust and repeatable method of approaching SLRs (Javed, 2019, Madurapperumage, 2021, Pattnaik, 2023). The structured approach makes it the most suitable literature review for policy recommendations and analysing multi-disciplinary spaces.

Comprehensive data retrieval (Bar-Ilan, 2018) is critical for research assessment. Therefore, four databases, Scopus, Web of Science (WoS), ACM DL, and IEEE Xplore, were selected to provide good coverage within our SLR.

So far, most of the existing academic work on digital technologies applied to sustainability has been published in technical journals. The ACM DL and IEEE are the most comprehensive and high-quality publishers within the technology domains and were selected for inclusion. Scopus and WoS are comprehensive interdisciplinary databases. In addition to covering Elsevier journals, other well-known academic publishers are also indexed in Scopus, including Springer, John Wiley and Sons. Due to the nature of the research question focusing on PB and the focal point of many of the blockchain solutions impacting emerging markets, ensuring the SLR included them as much as possible was deemed appropriate. Scopus has a solid global coverage, so it was considered essential to include. While including several databases created several duplicates, it was deemed suitable to use four to ensure that a broad range of peer-reviewed articles covering the interaction of digital technologies and PB were captured. These databases provide a relatively complete overview of high-quality literature in this space, which makes them suitable for literature analysis.

## Search Criteria

"Planetary Boundary" and "Digital" were selected as the primary keywords for this SLR. However, these terms can be used by authors in different ways. We, therefore, defined the following search criteria outlined in Table 1:

| Planetary Boundary(ies) | Digital Technologies |
| Planetary Boundary(ies) | Digital |
| Planetary Boundary(ies) | Digitalization/Digitalisation |

Table 1: Search Criteria

The same query was used for all four databases within the metadata – paper title, abstract and keywords. The search period was from January 1st, 1960, to August 31st, 2024. However, the seminal article outlining the concepts of Planetary Boundaries was published in 2009, so the effective search period is from 2009 to 2024. Only journal articles, conference papers and early-access articles were included in the search. 1,563 articles were obtained in total. Data downloaded included titles, authors, journal sources, abstracts, and references. 643 were duplicates, leaving 920 for initial review.

There were many duplicates due to the need to search SCOPUS and IEEE Access. However, removing the duplicates was considered preferable to missing any possible articles that SCOPUS had not indexed.

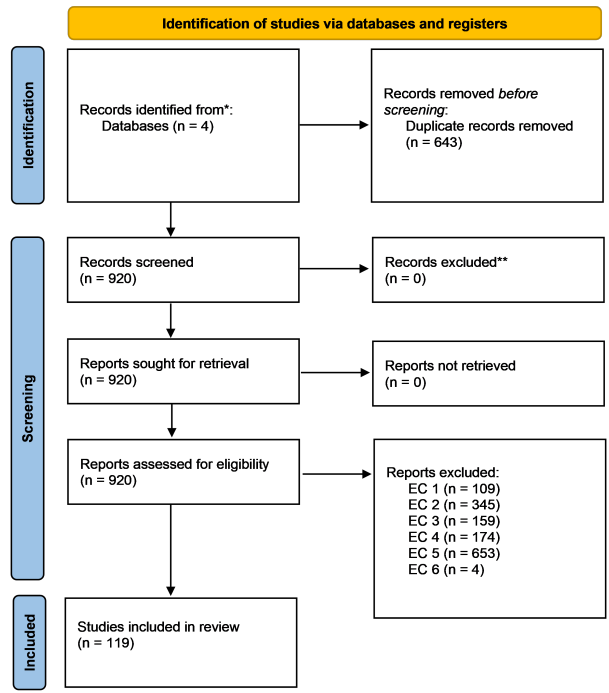

Figure 2: PRISMA Flowchart

## Evaluation Criteria

Eligibility assessment was conducted considering the following exclusion criteria:

- **EC 1:** Not peer-reviewed, e.g. book or book chapters, overviews of conferences and workshop proceedings (109 articles).
- **EC 2:** The work presented in the article does not qualify as research about Planetary Boundaries (345 articles) - for example, the key term "Planetary Boundary" refers to the Planetary Boundary Layer rather than the Planetary Boundaries concept.
- **EC 3:** Planetary Boundaries are not the main topic. Instead, it is used as a peripheral argument for discussing something else (159 articles).
- **EC 4:** Digital is mentioned only peripherally rather than as a driving solution in the paper. For example, the key term "digital" is mentioned as a supporting technology for teaching about Planetary Boundaries (174 articles).
- **EC 5:** The article has the same author(s), results, and methodological approach as another paper already included (653 articles).
- **EC 6:** Paper not in English (4 articles).

119 papers remained relevant to this paper's concepts used for the in-depth analysis. Literature was filtered according to the PRISMA approach. Data visualization was also applied to understand the overall research literature better. The open-source VOSviewer software was used for the analysis. Descriptive analysis focused on the spread of literature and which countries were involved. We then investigated the trends within the papers, identifying foci of the research and trends through clustering analysis of co-authors, co-keywords, and timelines of the papers. A PRISMA flowchart is shown in Figure 2.

# Results

Compared with papers on SDGs, there needs to be more focus on using digital technologies to help the world stay within PB. Overall, the publications were heavily weighted towards the ACM, reflecting the number of papers from the HCI community. It is interesting to note that few other publication outlets have seen as large a number of publications; this is, therefore, possibly a fruitful avenue for future research by the digital communities.  In addition, the number of papers increased sharply in 2022, but has receded somewhat – reflecting perhaps the complexity of solving PB with digital technologies, rather than claiming to have solved an SDG.

| Title | Number |
|---|---|
| AOM | 1 |
| Annual Reviews | 2 |
| Taylor and Francis | 2 |
| Nature | 6 |
| Elsevier | 10 |
| IEEE | 10 |
| ACM | 65 |

**Table 2: Spread of Publications**

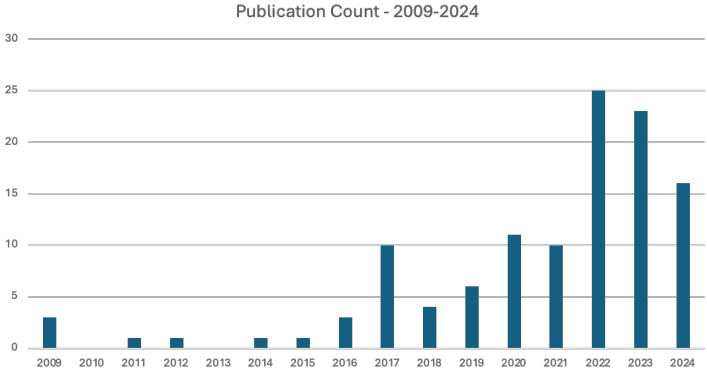

**Table 3: Publication Count 2009 - 2024**

Data visualisation was applied to investigate the literature. Three main research themes become clear can be seen through different aspects of visualisation:

1) Measurement: Many papers discuss using digital technologies to appropriately monitor and measure the shifts in boundaries in the PB approach (Danabasoglu et al, 2020, Paola et al, 2023)

2) Design: Several papers come from Human-Computer Interaction and are, as such, often speculative and conceptual. While there is value in such approaches, there are few concrete recommendations about how to build solutions, e.g (Stead et al, 2022, Akama et al, 2020, Kiourtis et al, 2024).

3)Circular Economy: Several papers link the concepts of digital technologies to generate solutions for the circular economy (Pouri and Hilty, 2020; Creutzig et al., 2022).

Investigating the timeline of the publications and the heatmaps illustrated that the focus on measurement and sustainability has continued but has also expanded over time to include the circular economy. An exciting contrast to the SDGs is that there is little focus on specific technologies such as IoT, Cloud or blockchain. This is likely reflective of the fact that, in contrast to the SDGs, it is required to take a complex systems view when using the PB approach hence the work is a little more complicated to fit within a normal scientific paper.

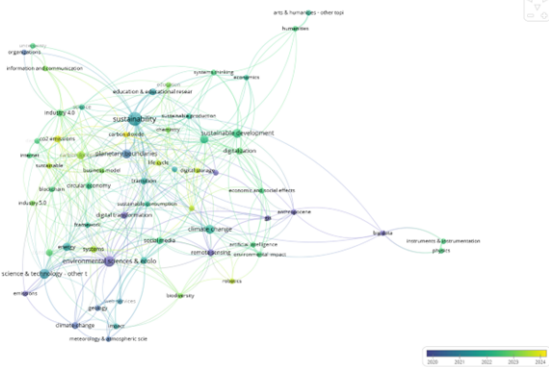

Figure 3: Timeline of Papers

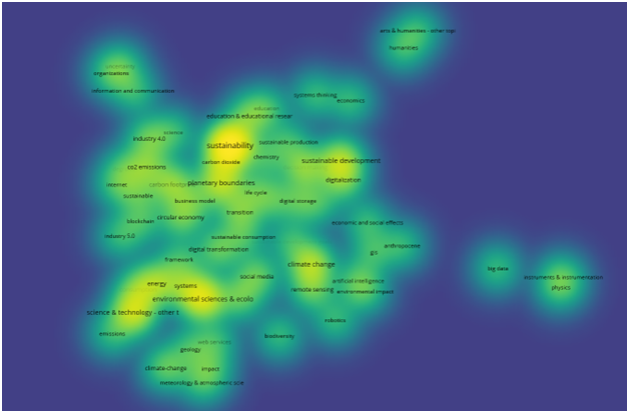

Figure 4: Heatmap of papers

A summary of the keywords' occurrences was created to refine the classifications further. Table 3 outlines the top 40 keywords, reinforcing the findings of the three main clusters.

| Keyword | Count |
|---|---:|
| Sustainability | 11 |
| Environmental Sciences and Ecology | 7 |
| Planetary Boundaries | 5 |
| Sustainable Development | 5 |
| Carbon Footprint | 4 |
| Climate Change | 5 |
| Decision Making | 4 |
| Energy | 4 |
| Science and Technology | 6 |
| Consumption | 3 |
| Remote Sensing | 3 |
| Transition | 3 |
| Industry 4.0 | 3 |
| Digitalisation | 3 |
| Education and Educational Research | 3 |
| Social Media | 3 |
| CO2 Emissons | 3 |
| Systems | 3 |
| Climate-Change | 3 |
| Modeling | 3 |
| Artificial Intelligence | 2 |
| Blockchain | 2 |
| Carbon Dioxide | 2 |
| Circular Economy | 3 |
| Digital Storage | 2 |
| Digital Transformation | 3 |
| Environmental Impact | 2 |
| Impact | 2 |
| Internet | 2 |
| Life-cycle | 2 |
| Sustainable Consumption | 2 |
| Sustainable Production | 2 |
| Biodiversity | 2 |
| Business Model | 2 |
| Chemistry | 2 |
| Framework | 2 |
| Information and Communication Technology | 2 |
| Instruments and Instrumentation | 2 |
| Physics | 2 |
| Web Services | 2 |
| Anthropocene | 2 |
| Big Data | 2 |
| Economic and Social Effects | 2 |
| Economics | 2 |
| Education | 2 |
| Emissions | 2 |
| Geology | 2 |
| GIS | 2 |
| Humanities | 2 |
| Information | 2 |
| Meteorology and Atmospheric Sciences | 2 |
| Science | 2 |
| Sustainable | 2 |
| Sustainable Development Goals | 2 |
| Degrowth | 2 |
| Systems Thinking | 2 |
| Industry 5.0 | 2 |
| Arts and Humanities | 2 |
| Organisations | 2 |
| Uncertainty | 2 |
| Machine Learning | 2 |
| Robotics | 2 |

Overall, within the SLR, no papers, however, cover an in-depth approach to PB and apply digital technologies to reframe how society thinks about delivering CNI. This is a critical gap, and we will provide some guidance around it in the following sections of the paper.

## Discussion: Gaps in Literature Exploiting Digital Capabilities for Planetary Boundaries

Rather than using digital technologies to reinforce our critical national infrastructures' long-standing organisational structures and design modalities, we should apply digital solutions to enable a modular approach to delivering essential services such as food, energy, and clean water supplies. Such a modular approach allows services to be delivered through dynamic strategic networks that function together as part of a larger supply chain. An added benefit of this approach is the flow-on effects of greater employment for local areas in different nations. This can go some way to reducing poverty, alongside other interventions (FAO, 2024)

However, in the case of a disruption, these dynamic strategic networks can reconfigure to provide an entire supply chain if required. In this instance, local communities already know how to work together to create resilience in delivering the services usually provided by CNI.

Digital technologies are applied to assist in smaller modules in the network:

- Co-ordinate in a timely manner regarding the products or services that are required by the larger supply chain (**Normal Operation**)
- Smooth supply to larger networks, enabling the stable delivery of required quantities and quality during stable periods (**Normal Operation**)
- Co-ordinate supply and demand side responses in unstable periods (**Resilience Operation**)

This concept is illustrated in Figures 7 and 8. The dynamic networks produce a combined product in X as input to a larger value chain. By using dynamic networks, smaller-scale suppliers can collaborate.

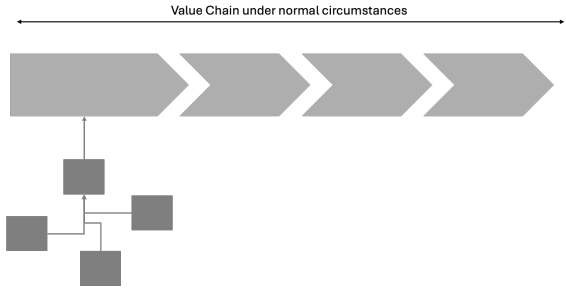

Figure 7: Dynamic Strategic Network enabled by digital technologies under normal conditions

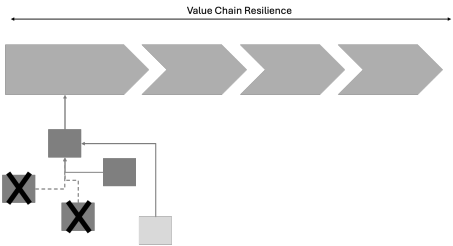

Figure 8: Dynamic Strategic Network enabled by digital technologies under resilience conditions

Through this approach, nations can deliver "Total Resilience" – a resilient nation that enables continued operations under large-scale climate change or other geopolitically induced problems in supply chains. It is essential to highlight that this approach is enabled by many different technologies, not just one. For example, many may suggest that such coordination is done using blockchain, while others may suggest IoT or cloud computing. However, in complex scenarios that include increasing climate events, relying on less complicated technologies may be more appropriate; we outline one example in the section below. The key to note here is that all digital technologies offer new methods to coordinate social and economic systems. By rethinking these systems from a PB lens, we can create new organisational structures that fit the 21st century and enable humanity to continue some semblance of normality while ensuring we live within our planet's boundaries. To fully explain this, the following section outlines the concept applied to food and agriculture.

## Application to Food and Agriculture

Food and agriculture critically impact both planetary boundaries and human existence. The food industry is also one of the clearest examples of 'brittle' supply chains. Large-scale supply chains have been constructed in - and across- most nations. Foodstuffs are grown in large-scale agricultural formats and are increasingly at risk of climate change through droughts, floods and biodiversity loss. Perhaps more than any other CNI, there is a close link between food production/retail and planetary boundaries. As early as 2012, it was clear that the world was: "in transition from an era of food abundance to one of scarcity…ushering in a new geopolitics of food. Food is the new oil. Land is the new gold" (Lester Brown). Climate change, geopolitical shifts and extreme weather events have contributed to an increasingly unstable food supply:

*"Food insecurity and malnutrition are worsening due to a combination of factors, including persisting food price inflation that continues to erode economic gains for many people in many countries. Major drivers like conflict, climate change, and economic downturns are becoming more frequent and severe. These issues, along with underlying factors such as unaffordable healthy diets, unhealthy food environments and persistent inequality, are now coinciding simultaneously, amplifying their individual effects."* (WHO, 2024).

Food and agriculture are cornerstones of human civilisation. Over several thousand years, many human civilisations have adapted their social structures to reflect the need to cultivate crops for food sources. Population increases have been met with adaptions to farming practices that increased crop yields. The world faces growing populations, limited crop yields, and large amounts of waste within the food supply chain and from consumers in Western economies. Previous solutions to these issues involved increasing agricultural productivity. Until recently, the global food industry has been highly focused on the supply side of food – ensuring that enough is produced and delivered to various regions of the world, i.e., food availability.

Over time, however, the approach has led to an extension of monocultures, a significant loss of agrobiodiversity, and accelerated soil erosion. The overuse of chemical fertilisers has polluted fresh water, increasing its phosphorus content and leading to a flow of phosphorus to the oceans, estimated to have risen to approximately 10 million metric tons annually. These human interventions in the natural environment are creating and combining with the complex interactions of climate change and seriously constrain the potential productivity of current agricultural methods.

By 2050, the world's population is estimated to reach 10 billion, with an estimated 56% shortfall in crop calories between those produced in 2010 and 2050 (WRI, 2024). In contrast with previous generations, merely increasing the production of existing crops alone is unlikely to achieve overall food security. Instead, by 2050, the food supply system must develop an ecological public health framework that balances food availability and health and nutrition requirements without compromising the planet's natural resources (Lang, 2009)

## Environmental Pressures

Environmental pressures on farming are three-fold: Firstly, monoculture farming reduces ecological diversity; secondly, depletion of fossil hydrocarbons will increase the demand for biofuels and industrial materials, which can create competition between food and biomass; finally, as natural resources are being depleted; climate change is of increasing importance. Agriculture is the single largest user of water worldwide, dwarfing everything else. It is calculated that it takes 15,000 litres of water to produce 1 kg of beef or 8,000 litres of water to produce a pair of jeans (Hoekstra, 2013). Industrial agriculture's water use is a cycle of over-use, waste and pollution. Approximately 3.8 trillion cubic meters of water are used by humans annually, with 70% consumed by the global agriculture sector (Worldometers, n.d).

Global meat production is expected to rise from 229 million metric tons between 1999 and 2001 to 465 million metric tons in 2050. This trend will put additional pressure on land and water systems, as more land and water are needed to produce meat than plant-based products of the same nutritional value (FAO, 2012)

## Sustainable and Healthy Diets

At the same time, concerns about the ability to provide healthy and environmentally sustainable diets for all people – are increasing. Sustainable food security, however, is not solely a problem of supply but a function of sustainable food sources and sustainable diets or "diets with low environmental impacts which contribute to food and nutrition security and healthy life for present and future generations. Sustainable diets are protective and respectful of biodiversity and ecosystems, culturally acceptable, accessible, economically fair and affordable; nutritionally adequate, safe and healthy; while optimising natural and human resources."(FAO, 2024)

As demonstrated by Stehfest et al (2009), a change in dietary patterns will have a dramatic impact on the environment. The authors calculate that a healthy diet worldwide would reduce the required area of arable land globally by 10%, the area of grassland by 40%, and the associated reduction in costs for mitigation of carbon dioxide emissions could be as much as 50% in 2050.

While dietary patterns influence the shape of food supply chains (FSCs), FSCs also determine the ability of consumers to make food choices. Consumers can only choose from the possibilities made available by networks of farms, manufacturers, retailers and businesses. The democratic access to health-enhancing diets is mediated by the structure and organisation of the FSC – and by price, income, class, location and culture (Lang, 2009).

## Complex Interacting Systems

Today, the food system is a net user of energy in virtually every nation; this is especially so in industrialized countries, where each calorie of food energy produced and brought to the table represents an average investment of about 7.3 calories of energy inputs (Heinberg, 2009) Declining fuel stocks are therefore also an issue for the food system. The FSC rests on an unstable foundation of massive fossil fuel inputs. Production of fertilisers, herbicides, pesticides, tillage, irrigation, fertilisation, transport, packaging, food conservation, and distribution worldwide require considerable energy.

The food industry faces a so-called bioenergy dilemma, in which difficult decisions must be made regarding whether crops are used to produce food or fuel. Waste is another boundary – as in other industries – but has potentially higher relevance in the food system as about 870 million people still suffer from chronic hunger globally (FAO 2024). The "nutrition transition" is a boundary related to demand that should be internalized in the supply side of food systems. The massive diversion of grain for fuel production has helped drive food prices up, leaving low-income consumers everywhere to suffer some of the most severe food price inflation in history. Food security is, therefore, a complex issue with multiple environmental, social, political, and economic determinants.

## Two Technology Choices

The world now faces a choice about how to apply technological innovation to the food supply system: to create a global "post-natural food system" in which technologies are used to increase yields or create new forms of bio-technical food systems, or to create a series of federated local/regional "renaturalised food systems" in which smaller scale local food systems are designed and supported through technology. Each approach must apply digital technologies to balance the needs of growing populations and scarce natural resources with food security.

1. In a post-natural world, technology is applied to push further beyond traditional agricultural practices and is more deeply embedded into the food system. In this scenario, ICT is applied beyond coordination technologies and becomes integrated as a core part of the production process. (**Traditional approach often seen in SDG applications**)
2. 2. A re-territorialized world, where technology supports resource conservation across the supply chain from supply to demand, focusing on agroecology. From this perspective, ICT is used to help coordinate the activities of multiple suppliers in dynamic strategic networks to coordinate the local delivery of food. ICT is required to successfully combine secure sustainable food systems through innovative digital technologies that consider the interactions between food and agriculture systems with broader industrial systems (**PB approach**)

Over time, a hybrid version of these two aspects will likely need to be developed, ensuring that the world can benefit from both approaches. This is illustrated in Figure 9.

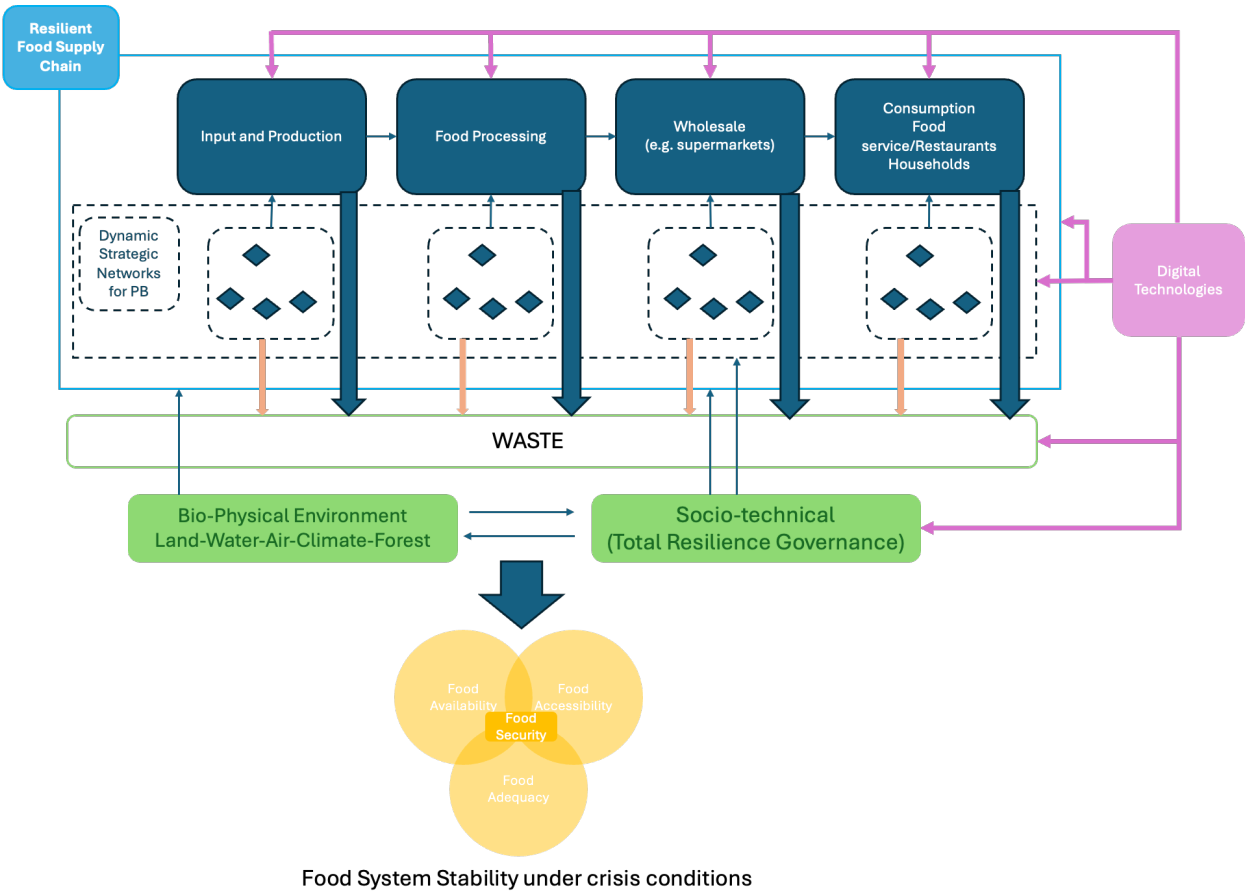

Figure 9: Illustration of digital technologies' role in delivering a resilient and stable food system

1. **Post-Natural World**

In the first scenario, digital technologies are used to increase the productivity of agricultural practices worldwide. For example, many farms are applying big data to improve operational analysis in the production of monocultures including maintenance of farming equipment, mapping of fields, and other operational activities to optimise watering and irrigation, the sowing of seeds, etc. These solutions are becoming economically viable due to the reduced cost of tailor-made sensor solutions, the lower price of storage and processing in cloud infrastructures and relatively cheap bandwidth (fixed and wireless) that permits the transmission of data sets from fields across nations and regions. The data is used not just to improve the performance of individual farms but – through data aggregation – to enhance performance for a significantly more significant number of farms purchasing seeds and equipment from the same supplier. Moreover, such data can be used in research and development to improve trials and create much larger data sets for analysis to develop new agricultural products, from GMO seeds to highly granular mapping software for farmers that permits companies to 'sell' optimal information for crop maintenance. These so-called "AgInformatics" systems are being heavily invested in by companies such as Deere Co, Dow AgroSciences and Monsanto, often reducing the farmers' agency and re-enforcing agricultural methods that increase the negative impacts of climate change.

Regarding resilience, however, these large-scale supply chains do not offer dynamic capabilities to re-organise if one or multiple parts are affected. The post-natural perspective may help feed more people, but it will not feed the extra 3 million people predicted, and more importantly, it will not enable a resilient food supply chain. If one part of the chain falls over, many will be left without sufficient foodstuffs, as was briefly illustrated in the COVID-19 pandemic.

2. **Re-Territorialized Food Supply Chains**

While a post-natural world will rely heavily on ICT to feed 10 billion people, another approach can be taken through the PB approach outlined in the previous section. A territorial food supply system comprises multiple 'short' supply chains organised in dynamic networks. This form of food system requires increased interaction with and response to the behaviours, preferences, and agricultural practices within different local areas, which cover rural, peri-urban, and urban spaces, as illustrated in Figure 10. To create these dynamic networks, each of the participants in the short supply chain must use and coordinate via digital technologies. Rural farms require long-haul logistics for their take-to-market activities. Peri-urban farms need regional medium-range coordination activities to deliver products to the correct market and customers. Within urban regions, another form of coordination is required to enable local logistics and coordination with more granular interactions to ensure that food arrives at the right time and place, in the correct quantity at the best price.

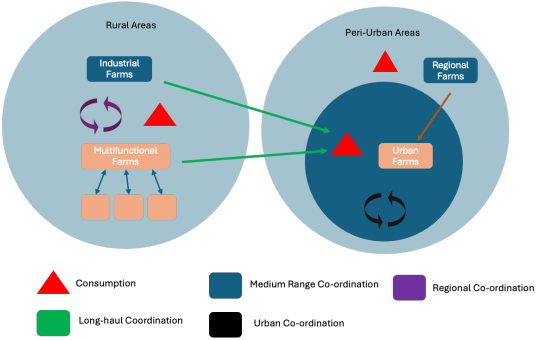

Figure 10: A Digitally-enabled resilient food supply chain using dynamic strategic networks

While the organisational strategy behind a short supply chain will differ from that of an industrialised food supply chain, the overall business processes remain the same. One significant difference between the food and industrial systems is that consumption needs to occur across all locations rather than mainly in urban or peri-urban areas, as may be the case for other physical products. Although the distribution, consumption and market preferences may differ, a consistent food supply is in equal demand across all areas – urban and rural. Within all the coordination scenarios in the creation of short supply chains, there is a strong dependence on the ability of digital technologies to arrange and effectively coordinate business processes across individuals who are not necessarily contractually obliged to one another. These strategic networks enable people to participate in a short supply chain when they can – e.g. when they have time or spare product available – but do not obligate them to participate. Through establishing networks with many other like-minded individuals, they can coordinate to ensure supply for customers from various regional and local suppliers.

## Conclusions

As the world begins to experience increasing climate change, nations must shift towards Total Resilience – where all aspects of society are tailored towards resilience rather than relying on large-scale brittle supply chains. This includes the creation of short supply chains that incorporate digitally enabled dynamic strategic networks. These networks can then best respond to the shifting needs on the ground. This paper has outlined the role of digital technologies in the delivery of these networks, with a focus on food and agriculture. The authors believe there are many similar organisational patterns in other supply chains – for example, energy and communications that could benefit from a similar organisational structure to deliver flexible, responsive and resilient CNI. This is a clear avenue of future research for those interested in applying digital technologies to create sustainability and resilience.

## References

Arjen Y. Hoekstra. (2013). *The water footprint of modern consumer society*. Routledge.

European Commission. (2024). *Consequences of climate change*. https://climate.ec.europa.eu/climate-change/consequences-climate-change_en

Food and Agriculture Organization (FAO). (2012). *Greening the economy with agriculture*.

Food and Agriculture Organization (FAO). (2024a). *Sustainable diets and biodiversity: Directions and solutions for policy, research and action*.

Food and Agriculture Organization (FAO), IFAD, UNICEF, WFP, & WHO. (2024b). *The state of food security and nutrition in the world 2024 – Financing to end hunger, food insecurity and malnutrition in all its forms*. https://doi.org/10.4060/cd1254en

Heinberg, R., & Bomford, M. (2009). *The food and farming transition: Toward a post-carbon food system*. Post Carbon Institute.

Javed, M. N., Shafiq, H., Alam, K. A., Jamil, A., & Sattar, M. U. (2019). VANET's security concerns and solutions: A systematic literature review. *Proceedings of the 2019 International Conference on Advanced Information Science and System (AISS)*, Article 40. https://doi.org/10.1145/3341325.3342028

Kitchenham, B., & Charters, S. (2007). *Guidelines for performing systematic literature reviews in software engineering*(EBSE Technical Report EBSE-2007-01). https://www.cs.auckland.ac.nz/~norsaremah/2007%20Guidelines%20for%20performing%20SLR%20in%20SE%20v2.3.pdf

Madurapperumage, A., Wang, W. Y. C., & Michael, M. (2021). A systematic review on extracting predictors for forecasting complications of diabetes mellitus. *Proceedings of the 2021 International Conference on Advanced Information Science and System (AISS)*, 327–330. https://doi.org/10.1145/3472813.3473211

Page, M. J., Moher, D., Bossuyt, P. M., Boutron, I., Hoffmann, T. C., Mulrow, C. D., ... & McKenzie, J. E. (2021). PRISMA 2020 explanation and elaboration: Updated guidance and exemplars for reporting systematic reviews. *BMJ*, 372, n160. https://doi.org/10.1136/bmj.n160

Pattnaik, N., Li, S., & Nurse, J. (2023). A survey of user perspectives on security and privacy in a home networking environment. *ACM Computing Surveys*, 55(9), 38. https://doi.org/10.1145/3558095

Stead, M., Coulton, P., Pilling, F., Gradinar, A., Pilling, M., & Forrester, I. (2022). More-than-human-data interaction: Bridging novel design research approaches to materialise and foreground data sustainability. *Proceedings of the 2022 ACM Conference on Human Factors in Computing Systems (CHI)*, 62–74. https://doi.org/10.1145/3569219.3569344

Stehfest, E., Bouwman, L., van Vuuren, D. P., & others. (2009). Climate benefits of changing diet. *Climatic Change*, 95, 83–102. https://doi.org/10.1007/s10584-008-9534-6

World Health Organization. (2024, July 24). *Hunger numbers stubbornly high for three consecutive years as global crises deepen: UN report*. https://www.who.int/news/item/24-07-2024-hunger-numbers-stubbornly-high-for-three-consecutive-years-as-global-crises-deepen--un-report

World Resources Institute. (2018, December 5). *How to sustainably feed 10 billion people by 2050, in 21 charts*. World Resources Institute. https://www.wri.org/insights/how-sustainably-feed-10-billion-people-2050-21-charts

Worldometers. (n.d.). *Water usage statistics*. Worldometers. Retrieved September 29, 2024, from https://www.worldometers.info/water/

## Competing interests (mandatory)

The authors have no competing interests to declare

## Acknowledgements (optional)

Parts of this research were paid for by EPSRC grant: EP/J000604/2 and Ericsson Networked Society Lab

## Author contributions

C.M. conceived the experiment(s), G.B. and S.G. conducted the experiment(s), C.M. G.B. and S.G. analysed the results.  All authors reviewed the manuscript.